\documentclass[12pt,preprint]{emulateapj}

\newcommand{\be}{\begin{equation}}
\newcommand{\ee}{\end{equation}}
\newcommand{\ba}{\begin{eqnarray}}
\newcommand{\ea}{\end{eqnarray}}

\begin{document}
\title{Axion decay and anisotropy of near-IR extragalactic background light}

\author{Yan Gong$^{1,2}$, Asantha Cooray$^2$, Ketron Mitchell-Wynne$^2$, Xuelei Chen$^{1,3}$, Michael Zemcov$^4$ and Joseph Smidt$^5$}

\affil{$^{1}$ National Astronomical Observatories, Chinese Academy of Sciences, Beijing, 100012, China \\
$^2$Department of Physics \& Astronomy, University of California, Irvine, CA 92697 \\
$^3$ Center of High Energy Physics, Peking University, Beijing 100871, China\\
$^4$ Center for Detectors, School of Physics and Astronomy, Rochester Institute of Technology, Rochester, NY 14623, USA\\
$^5$ Theoretical Division, Los Alamos National Laboratory, Los Alamos, New Mexico 87545, USA\\
}

\begin{abstract}
The extragalactic background light (EBL) is comprised of the cumulative radiation from all galaxies and active galactic nuclei over the cosmic history. In addition to point sources, EBL also contains information from diffuse sources of radiation. The angular power spectra of the near-infrared intensities could contain additional signals and a complete understanding of the nature of the IR background is still lacking in the literature. Here we explore the constraints that can be placed on particle decays, especially candidate dark matter models involving axions that trace dark matter halos of galaxies. Axions with a mass around a few eV will decay via two photons with wavelengths in the near-IR band, and will leave a signature in the IR background intensity power spectrum. Using recent power spectra measurements from the Hubble Space Telescope (HST) and Cosmic Infrared Background Experiment (CIBER), we find that the 0.6 to 1.6 micron power spectra can be explained by axions with masses around 4 eV. The total axion abundance $\Omega_a\simeq0.05$, and it is comparable to the baryon density of the Universe. The suggested mean axion mass and abundance are not ruled out by existing cosmological observations. Interestingly, the axion model with a mass distribution is preferred by the data, which cannot be explained by the standard quantum chromodynamics (QCD) theory and needs further discussion.
\end{abstract}


\maketitle

\section{Introduction}

The extragalactic background light (EBL) from ultraviolet (UV) to far-infrared (far-IR) bands contains the cumulative radiation emitted from all galaxies  and active galactic nuclei (AGNs) over the cosmic history. There are two peaks on the EBL intensity spectrum located at near-infrared and far-infrared bands, which are expected to be contributed by direct emission from stars in galaxies and re-emission from dust heated by UV/optical photons, respectivel \citep{Baldry03,Fukugita04,Franceschini08,Finke10, Stecker12, Inoue13,Gong13}. At near-IR wavelengths there is a 
large discrepancy between measurements of absolute EBL intensity and the expectations from models of galaxy
formation and evolution, in that the absolute measurements are higher than the galaxy counts by at least an order of magnitude at a wavelength around 1 $\mu$m.  This difference is likely due to large systematic uncertainties in the absolute measurements involving the foreground contamination from Zodiacal light, and to a lesser extent from our Galaxy \citep{Hauser01}. 

Instead of absolute intensity, the recent measurements focus on the anisotropies of the background. This is due to the fact that the foregrounds such as Zodiacal light have smooth spatial distributions and have a different correlation function than the fluctuations generated by extragalactic signals. The current measurements of the near-infrared intensity fluctuations suggest an excess in the clustering signal at scales of a few arcminutes, relative to galaxy clustering \citep{Kashlinsky05,Thompson07,Kashlinsky07,Matsumoto11,Kashlinsky12,Cooray12a,Cooray12b,
Zemcov14}. This has allowed new ideas involving diffuse sources, such as intra-halo light (IHL) \citep{Cooray12b,Zemcov14}, or direct collapse blackholes (DCBHs) from very high redshifts \citep{Yue13}. The existing measurements and model development is such that we still have room to explore additional signals. Here we focus on the possibility involving the decay products of axion particles that can also be considered as a viable candidate for dark matter.

The axion is a hypothetical particle that was proposed to solve the strong color parity (CP) problem \citep{Weinberg78,Wilczek78}. It is created during the breaking of the Peccei-Ouinn (PQ) symmetry at an energy scale that is yet to be determined experimentally, thus labeled as $f_a$ hereafter \citep{Peccei77}. It is also a viable candidate for cold dark matter. Axions decay into two photons with a lifetime that depends on the mass $m_a$ and the coupling strength $g_{a\gamma\gamma}$ \citep{Ressell91}. For $m_a\lesssim10^{-2}$ eV, axion is non-thermal, and its lifetime is much longer than the age of the Universe. In such a scenario axions are totally invisible. On the other hand, for $m_a>10^{-2}$ eV, axions can be produced by thermal mechanisms in the early Universe \citep{Turner87} and their presence is detectable in the multi-eV window \citep{Overduin04}.

In this paper we use axion-photon decay to interpret the existing measurements of intensity power spectra at optical to near-IR wavelengths. This in turn will constrain the properties of the eV-mass axions, including the allowed mass range  and coupling strength to photons. Although the thermal axion is not as cold as non-thermal axions, we assume here that they can and will be captured by the gravitational potential well primarily formed by another cold dark matter candidate, such as a weakly interacting massive particles (WIMPs). These halos are the locations of galaxies and galaxy clusters \citep{Kephart87}. The decay of axions in such dark matter halos will lead to a signal that is similar to IHL (Cooray et al. 2012b) in terms of the spatial structure. The differences will be in the redshift evolution of the signal. The fluctuation signal is such that it may explain the excess clustering that is seen in near-IR anisotropy power spectrum.

To compare our model predictions with the data and to constrain the overall model related to the presence of axions,
we make use of the latest intensity power spectra measurements from the {\it Hubble} Space Telescope (HST) and Cosmic Infrared Background Experiment (CIBER). These measurements lead to EBL anisotropies in seven optical and near-IR bands (five bands of HST and two bands of CIBER) from 0.6 to 1.65 $\mu$m \citep{Zemcov14,Mitchell-Wynne15}. Our overall model includes contributions from axion-photon decay, low-redshift ($z<6$), and high-redshift ($z>6$) galaxies. We assume that the suggested signals such as IHL and DCBHs are zero for the purposes of obtaining the maximally allowed axion abundance. In reality, with such signals present, the actual axion abundance may be lower. Given that we have large uncertainties on the exact level of IHL and DCBH signals, we do not have a reliable way to constrain the exact abundance here apart from an upper limit. The upper limit, however, as we discuss below, is still lower than the total energy density of dark matter suggesting that the axion of 4 eV mass scale alone cannot explain the total cold dark matter content of the universe. We make use of the Markov Chain Monte Carlo (MCMC) method to model-fit the data and to obtain constraints on the axion mass and coupling strength with photons. We assume the flat $\Lambda$CDM model with $\Omega_{M}=0.27$, $\Omega_{b}=0.046$, and $h=0.71$ for the calculation throughout the paper.

This paper is organized as follow: in Section 2, we discuss the details of our calculation to derive the luminosity, mean intensity, 
and angular power spectra of axion-photon decay; In Section 3, we show power spectrum data from HST and CIBER observations we use, 
and the details of the model we adopted in the MCMC fitting process; 
We show the fitting results in Section 4 and give a summary and discussion in Section 5.

\section{Mean intensity and anisotropies of Axion decay}

In this Section, we discuss the method that we use to estimate the mean intensity and anisotropies from the axion-photon decay. First, we need to estimate the luminosity of axion decay in a halo with mass $M$. The luminosity of a halo contributed by the axion-photon decay is given by \citep{Kephart87}
\be \label{eq:La}
L_a = \frac{M_a c^2}{\tau_a},
\ee
where $M_a$ is the total axion mass in the halo, and we assume the fraction of axion density in a halo is the same as the mean fraction in the Universe\footnote{We notice that, according to the generalized Tremaine-Gunn bound \citep{Madsen90,Madsen91,Ressell91}, the axion density is subject to the phase-space limits which can lead to smaller axion fraction in galaxy-mass halos than the mean fraction in the Universe. We find this suppression effect would not significantly affect our results, since the main contribution of axion-photon decay is from cluster-mass halos.}, which gives
\be \label{eq:Ma}
M_a = \frac{\Omega_a}{\Omega_M} M.
\ee
Here $M$ is the total halo mass, $\Omega_a=\rho_a/\rho_{\rm crit,0}$ is the present axion mean density parameter of the Universe, and $\rho_a$ and $\rho_{\rm crit,0}$ are the present axion mass density and critical density of the Universe. The $\Omega_M=\Omega_a + \Omega_c + \Omega_b$ is the total matter density parameter, and $\Omega_c$ and $\Omega_b$ are the mean density parameters for cold dark matter (e.g. WIMPs) and baryons, respectively. Since the axion mass range we consider is effectively thermally cold, its energy density evolves as $(1+z)^3$ which is the same as the cold dark matter and Eq.~(\ref{eq:Ma}) is accurate for all redshifts.

For axions with mass greater than $\sim 10^{-2}$ eV, they are produced thermally and are in thermal equilibrium with the other particle species in the early Universe \citep{Turner90}. These thermal axions decouple when they are still relativistic and their comoving number density is effectively ``frozen in" during the subsequent evolution of the universe \citep{Turner90}. Thus, the present-day density parameter of thermal axions $\Omega_a$ can be obtained by solving the Boltzmann equation. If the number of relativistic degrees of freedom when axions ``froze out'' is around 15, $\Omega_a$ is given by \citep{Turner90,Overduin04}
\be \label{eq:Omega_a}
\Omega_a = 5.2 \times 10^{-3} h^{-2}\,\frac{m_a}{\rm eV},
\ee
where $m_a$ is the axion particle mass in eV, and $h=H_0/(100\ \rm km\,s^{-1}Mpc^{-1})$ where $H_0$ is the Hubble constant today. We can find an upper limit for $m_a$ using Eq.(\ref{eq:Omega_a}) and assuming axions contribute all the dark mater density such that $\Omega_a=\Omega_M-\Omega_b$. This gives $m_a\lesssim22$ eV.

In Eq.(\ref{eq:La}), $\tau_a$ is the axion lifetime for the process of an axion decaying to a photon pair and is given by \citep{Ressell91}
\be \label{eq:taua}
\tau_a = (6.8 \times 10^{24}\ s) \left(\frac{m_a}{\rm eV} \right)^{-5} \zeta^{-2},
\ee
where $\zeta=|E/N-1.95|/0.72$ is the normalized coupling constant. Here $E$ and $N$ are the values of the electromagnetic anomaly and color anomaly of PQ symmetry, respectively. We have $\zeta=1$ for the simplest unified models of strong and electroweak interactions \citep{Ressell91}. The $\zeta$ is proportional to axion-photon coupling strength $g_{a\gamma\gamma}$, and the relation is
\be \label{eq:zeta_gamma}
g_{a\gamma\gamma} = 0.72\, \zeta \frac{\alpha_f}{2\pi f_a}.
\ee
Here $\alpha_f=1/137$ is the fine-structure constant, and $f_a=f_{\rm PQ}/N=6.2\times 10^6/(m_a/{\rm eV})$ GeV where $f_{\rm PQ}$  is the energy scale at which axions are created when U(1) PQ symmetry breaks down \citep{Ressell91}. In principle, $\zeta$ can be much smaller than 1 and  can even vanish completely in some stable axion models (Kaplan 1985; Ressell 1991; Overduin \& Wesson 2004). Here we take $0<\zeta\le 1$ as a prior range in our model fits described later. In Eq. (\ref{eq:taua}), we find $m_a\simeq 27.5$ eV if $\zeta=1$ and the age of the Universe $t_{\rm univ}\simeq 4.33\times10^{17} s$ as the axion lifetime. Of course, $m_a$ can be much smaller if $\tau_a$ is much larger than $t_{\rm univ}$. 

From Eq.(\ref{eq:La})-(\ref{eq:taua}), the luminosity contributed by the axion-photon decay in a halo with mass $M$ is
\be\label{eq:La_e} 
L_a(M) = 2.6 \times 10^{-6} \left(\frac{m_a}{\rm eV}\right)^6 \zeta^2\frac{M}{M_{\sun}}\ L_{\sun}.
\ee
Hence, approximately, axions with mass around 10 eV  with $\zeta\simeq 0.1$ can provide a luminosity  comparable to the galaxy emission in a halo, leading to $L_a\sim 10^{-2}(M/M_{\sun})$$L_{\sun}$ or mass-to-light ratios of 100 \citep{Kephart87}.  We also note that, according to Eq.~(\ref{eq:La_e}), $m_a$ and $\zeta$ are not sensitive to the fraction of axion in dark matter halos. This is because $L_a$ is proportional to the axion fraction in halos from Eq.~(\ref{eq:La}) and Eq.~(\ref{eq:Ma}), and we have $L_a\sim m_a^6 \zeta^2$ which indicates that $m_a\sim L_a^{1/6}$ and $\zeta\sim L_a^{1/2}$. With such low dependency, $m_a$ and $\zeta$ are mostly independent of axion fraction.

Here, instead of a single mass for axions, we also consider a mass distribution. Inspired by the string theory \citep[e.g.][]{Svrcek06,Arvanitaki10} and experiments of high energy physics, we assume a general form of axion mass distribution as
\be \label{eq:P_ma}
P(m_a) = \frac{\alpha}{m_0} \frac{1}{\Gamma(1+\frac{1}{\alpha})} \left( \frac{m_a}{m_0}\right)^{\alpha} {\rm exp}\left[ -\left( \frac{m_a}{m_0}\right)^{\alpha}\right],
\ee
where $m_0$ and $\alpha$ are free parameters, and $\Gamma(x)$ is the Gamma function. Note that this form is phenomenal, and our purpose is to derive the mass distribution from the observational data and minimize theoretical assumptions. We find this form is good enough for the constraints, and more general forms with more free parameters would not change the results. The mean mass of axion particles, under such a distribution, is
\be \label{eq:ma_m}
\bar{m}_a = \int m_aP(m_a) dm_a.
\ee
In this case, we calculate $\Omega_a$ and mean axion lifetime $\bar{\tau}_a$ by replacing $m_a$ with $\bar{m}_a$ in Eqs.~(\ref{eq:Omega_a}) and (\ref{eq:taua}).

To estimate the mean intensity of axion-photon decay at different wavelengths, we also need spectral energy distribution (SED) of the decay photons. Since each photon in the decay photon pair has the same energy $\frac{1}{2}m_a c^2$, the wavelength of each decay photon should be
\be \label{eq:l_ma}
\lambda_a = \frac{h_p c}{\frac{1}{2}m_a c^2} \simeq \frac{2.48}{m_a/\rm eV}\  \rm \mu m,
\ee
where $h_{\rm p}$ is the Planck constant. This wavelength is around near infrared and optical bands when $1\lesssim m_a\lesssim 10$ eV, which is the mass range of the axions that we will be interested in this paper.

For the case of a single axion mass, following Ressell (1991) and Overduin \& Wesson (2004), we use a Gaussian rest-frame SED to consider axions bounded in the galaxy cluster halo
\be \label{eq:F_l_ma}
F(\lambda) = \frac{1}{\sqrt{2\pi}\sigma_{\lambda}}\,{\rm exp}\left[-\frac{1}{2}\left( \frac{\lambda-\lambda_a}{\sigma_{\lambda}}\right)^2 \right],
\ee
where $\sigma_{\lambda}=2(v_a/c)\lambda_a$ is the standard deviation, which can be derived from the velocity dispersion of the axions $v_a$ in the halo. Note that $v_a$ is as a function of halo mass $M$, and we assume $v_a(M)\sim \sqrt{M}$ according to the general relation of velocity dispersion and halo mass. However, we find the intensity spectrum of axion-photon decay is insensitive to the value of $v_a$ unless $v_a$ is close to $10^4$ km s$^{-1}$. This is because the velocity dispersions for galaxies or galaxy clusters always provide relatively small $\sigma_{\lambda}$, and this results in narrow SED profiles and makes Eq.~(\ref{eq:F_l_ma}) approach a delta function. For instance, for large galaxy clusters with a mass of 10$^{15}$ M$_{\sun}$, we have $v_a=1300\ \rm km\,s^{-1}$ which leads to $\sigma_{\lambda}\simeq 220 {\rm \AA}/({m_a/\rm eV})$ (Ressell 1991; Overduin \& Wesson 2004). In the near-IR band with $\lambda\sim \mathcal{O}(1)$ $\mu$m, this $\sigma_{\lambda}$ is about two orders of magnitude smaller than $\lambda$ with $m_a\sim  \mathcal{O}(1)$ eV, and it would be smaller for galaxy clusters with smaller masses. Hence, for simplicity and considering the accuracy and efficiency of our calculation (especially in the MCMC process), we adopt an uniform $v_a$ for all halo masses. Our results with this assumption is identical to that with $v_a(M)$ for the accuracy required in this work.

For the case with an axion mass distribution, we find
\be
F(\lambda) = P(\lambda),
\ee
where $P(\lambda)$ is the normalized probability distribution for the wavelength $\lambda=\lambda_a$, which can be derived from $P(m_a)$  and $\lambda_a(m_a)$ given by Eqs.~(\ref{eq:P_ma}) and (\ref{eq:l_ma}). Note that we do not consider the velocity dispersion effect given by Eq.~(\ref{eq:F_l_ma}) for the case 
involving an axion mass distribution, since the dispersion at each halo mass is negligible 
compared to the dispersion coming from $P(\lambda)$ which has a much wider wavelength distribution.

Then the luminosity of axion decay at $\lambda$ is given by
\be
L_{a,\lambda} = L_a F(\lambda) = L_a(M)F[\lambda_0/(1+z)],
\ee
where $\lambda_0$ is the observed wavelength at $z=0$. Now we can estimate the comoving emissivity of axion-photon decay by
\be
\bar{j}_{\lambda}(z) = \frac{1}{4\pi}\int dM \frac{dn}{dM}\, L_{a,\lambda}(M,z).
\ee
Here ${dn}/{dM}(M,z)$ is the halo mass function (Cooray \& Sheth 2002). The mean intensity of axion-photon decay at observed wavelength $\lambda_0$ is given by
\be \label{eq:Il}
I_{\lambda}(\lambda_0) = \int_0^{z_{\rm max}} \frac{c\,dz}{H(z)(1+z)^3}\,\bar{j}_{\lambda}(z),
\ee
where $H(z)$ is the Hubble parameter, and we assume the flat $\Lambda$CDM model with $H(z)=\sqrt{\Omega_M(1+z)^3+\Omega_{\Lambda}}$ and $\Omega_{\Lambda}=1-\Omega_M$. The $z_{\rm max}$ is the maximum redshift we consider, and we take $z_{\rm max}=30$ here. Note that in the case of a single axion mass, 
$I_{\lambda}$ at different wavelengths originates from different redshifts, since the SED has a Gaussian profile with relatively small $\sigma_{\lambda}$. 
Since structure at different redshift intervals are mainly uncorrelated, this
suggests that the cross-correlation between intensity fluctuations at different wavelengths will be zero, contrary to the observations.
Our motivation for using a mass distribution is exactly because of this reason: in order to preserve the observed
cross-correlations between bands from 0.6 to 1.6 $\mu$m we must allow different axion masses to contribute at the same redshift.

The 1-halo and 2-halo terms of the angular cross power spectrum for axion-photon decay at observed wavelengths $\lambda_0$ and $\lambda'_0$ can be evaluated by
\ba
C_{\ell,\rm 1h}^{\lambda_0\lambda'_0} = &&\frac{1}{(4\pi)^2} \int dz \left( \frac{d\chi}{dz}\right) \left( \frac{a}{\chi}\right)^2 \nonumber \\
                                 &&\times \int dM \frac{dn}{dM} u^2(k|M,z) L_{a,\lambda} L_{a,\lambda'},
\ea
\ba
C_{\ell,\rm 2h}^{\lambda_0\lambda'_0} = &&\frac{1}{(4\pi)^2} \int dz \left( \frac{d\chi}{dz}\right) \left( \frac{a}{\chi}\right)^2 P_{\rm lin}(k,z) \nonumber \\
                                        &&\times \int dM \frac{dn}{dM} b(M,z) u(k|M,z) L_{a,\lambda} \nonumber \\
                                        &&\times \int dM \frac{dn}{dM} b(M,z) u(k|M,z) L_{a,\lambda'}.
\ea
Here $\chi$ is the comoving distance, $a=1/(1+z)$ is the scale factor, $b(M,z)$ is the halo bias,  and $P_{\rm lin}(k,z)$ is the linear power spectrum with $k=\ell/\chi$ (Cooray \& Sheth 2002). The $u(k|M,z)$ is the Fourier transform of axion density profile in a halo, and we assume it follows the Navarro-Frenk-White (NFW) profile (Navarro et al. 1997). The total angular cross power spectrum for $\lambda_0$ and $\lambda'_0$ is then given by
\be
C_{\ell}^{\lambda_0\lambda'_0} = C_{\ell,\rm 1h}^{\lambda_0\lambda'_0} + C_{\ell,\rm 2h}^{\lambda_0\lambda'_0}.
\ee
Then we use $C_{\ell}=\lambda_0\lambda'_0 C_{\ell}^{\lambda_0\lambda'_0}$ to fit the data from the observations as we show in the next section. We notice that the free-streaming effect of axion can suppress the angular power spectrum at small scales \citep[e.g.][]{Hannestad05,Hannestad10, Grin08}, but we find this effect would not change our constraints on $m_a$ and $\zeta$ significantly for $m_a< 5.5$ eV. Hence, we ignore this effect in the data-fitting process for simplicity.

\section{Model constraint}

We primarily make use of two datasets in our fitting process.
First is from the HST observations in five optical and near-IR bands centered at 0.606, 0.775, 0.850, 1.25 and 1.6 $\mu$m. The dataset is obtained by the Wide Field Camera 3 (WFC3) and Advanced Camera for Surveys (ACS),  and it covers 120 square arcminutes in the Great Observatories Origins Deep Survey (GOODS) \citep{Mitchell-Wynne15}. The second is from CIBER, a rocket-borne instrument that was designed for measuring the spatial and spectral properties of the EBL, with fluctuation measurements in two near-IR bands centered at 1.1 and 1.6 $\mu$m. The CIBER data are from two flights in 2010 and 2012, using two 11-cm telescopes each with a four square degree field of view \citep{Zemcov14}. We also included fluctuation data from {\it Spitzer} observations at 3.6 $\mu \rm m$ given by \cite{Cooray12b}, but found that cannot fit those measurements around $\ell=3\times10^3$ with an adequate chi-square value, especially for an axion model involving a single mass. We give details and a discussion related to {\it Spitzer} model fits in the Appendix. 

\begin{figure*}[t]
\epsscale{1.9}
\centerline{
\resizebox{!}{!}{\includegraphics[scale=0.45]{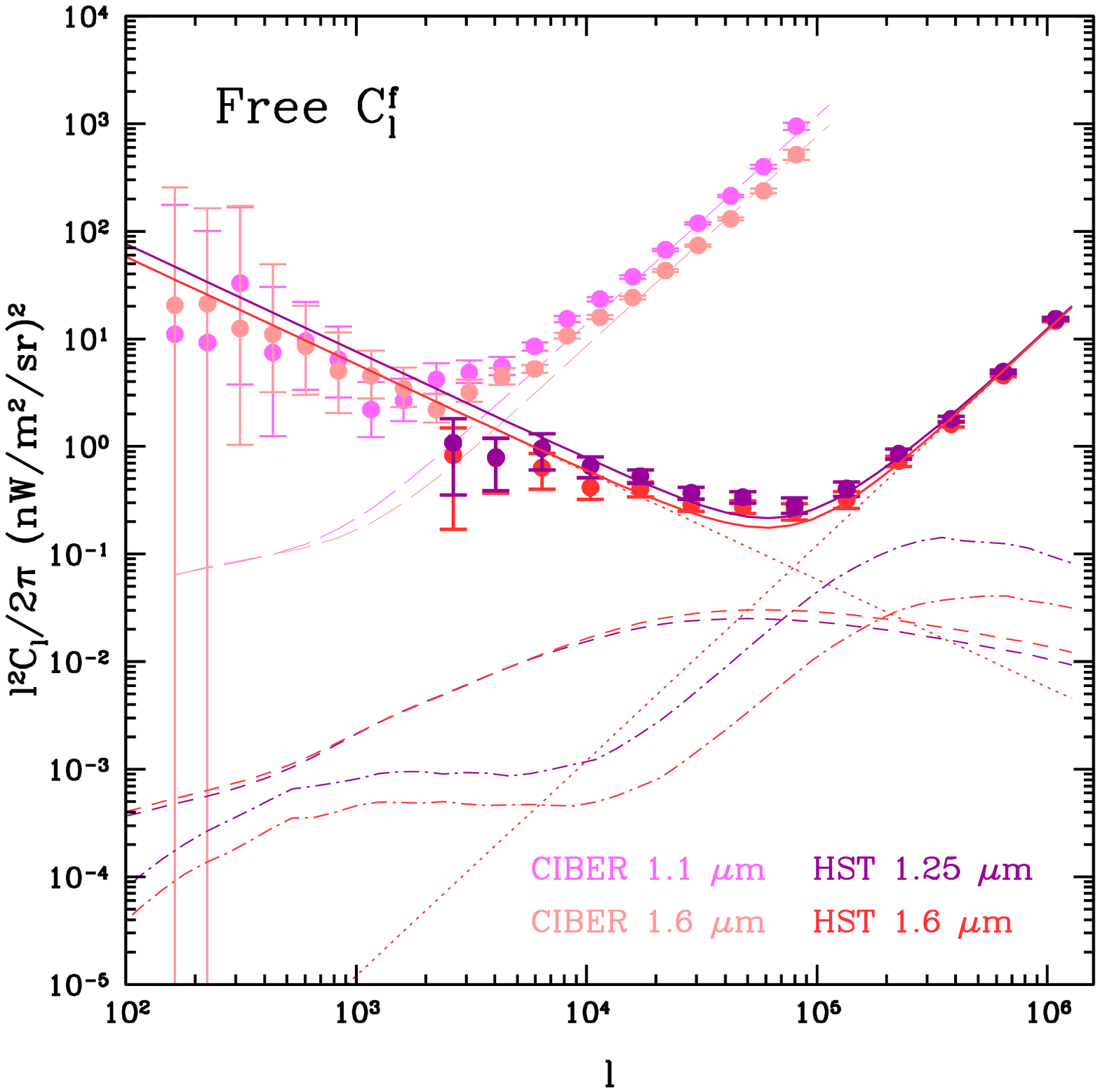}}
\resizebox{!}{!}{\includegraphics[scale=0.45]{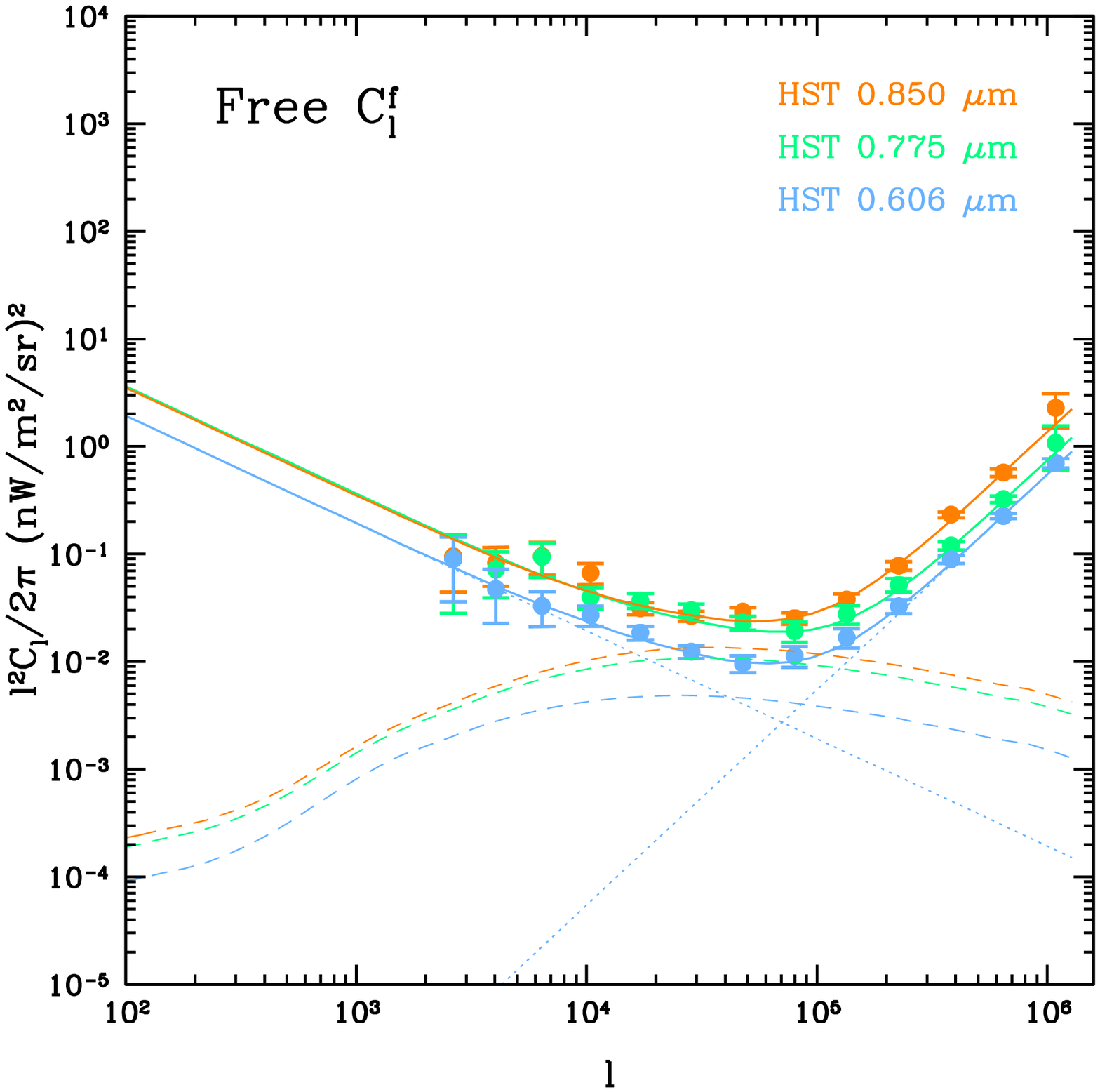}}
}
\centerline{
\resizebox{!}{!}{\includegraphics[scale=0.45]{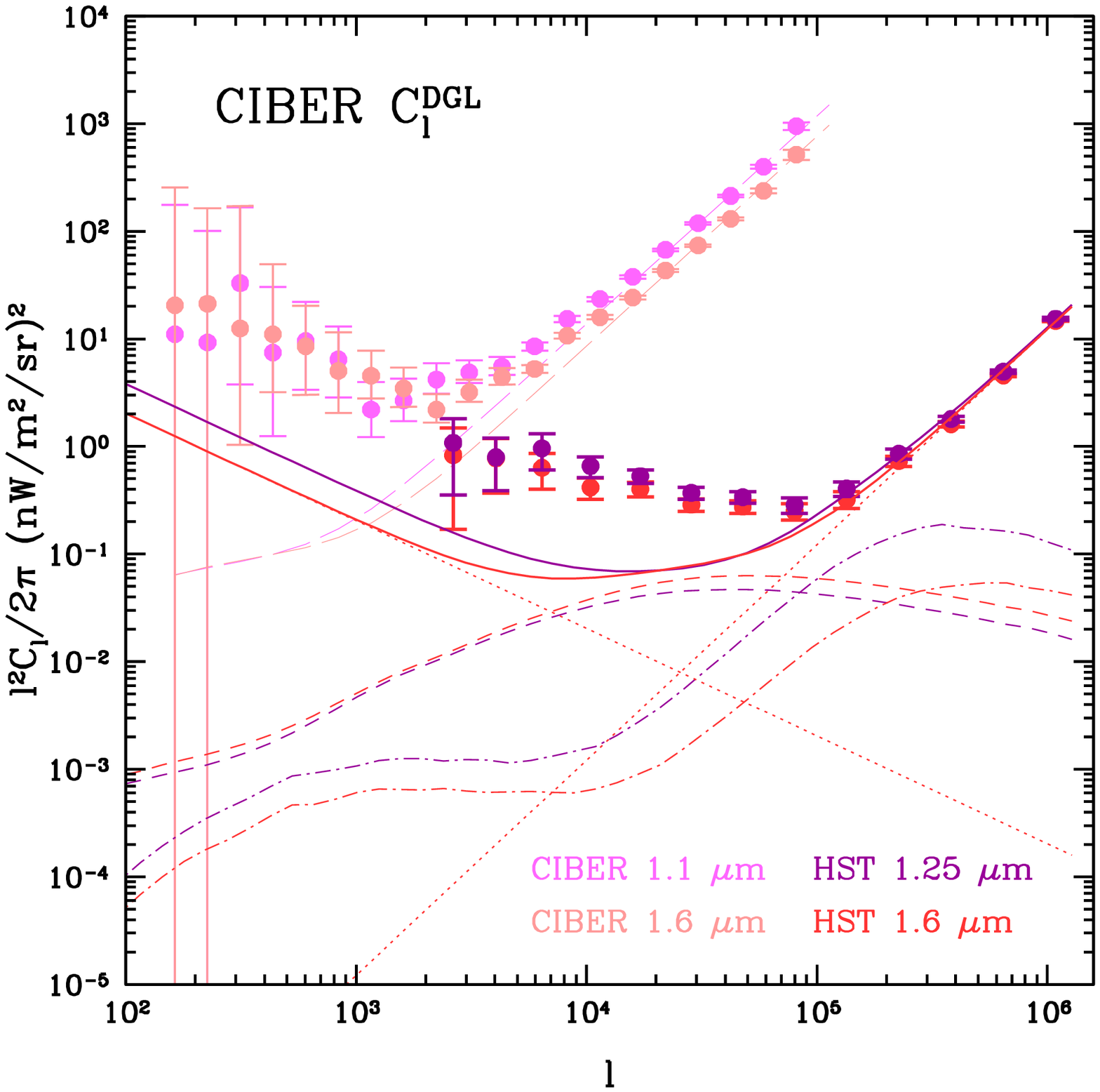}}
\resizebox{!}{!}{\includegraphics[scale=0.45]{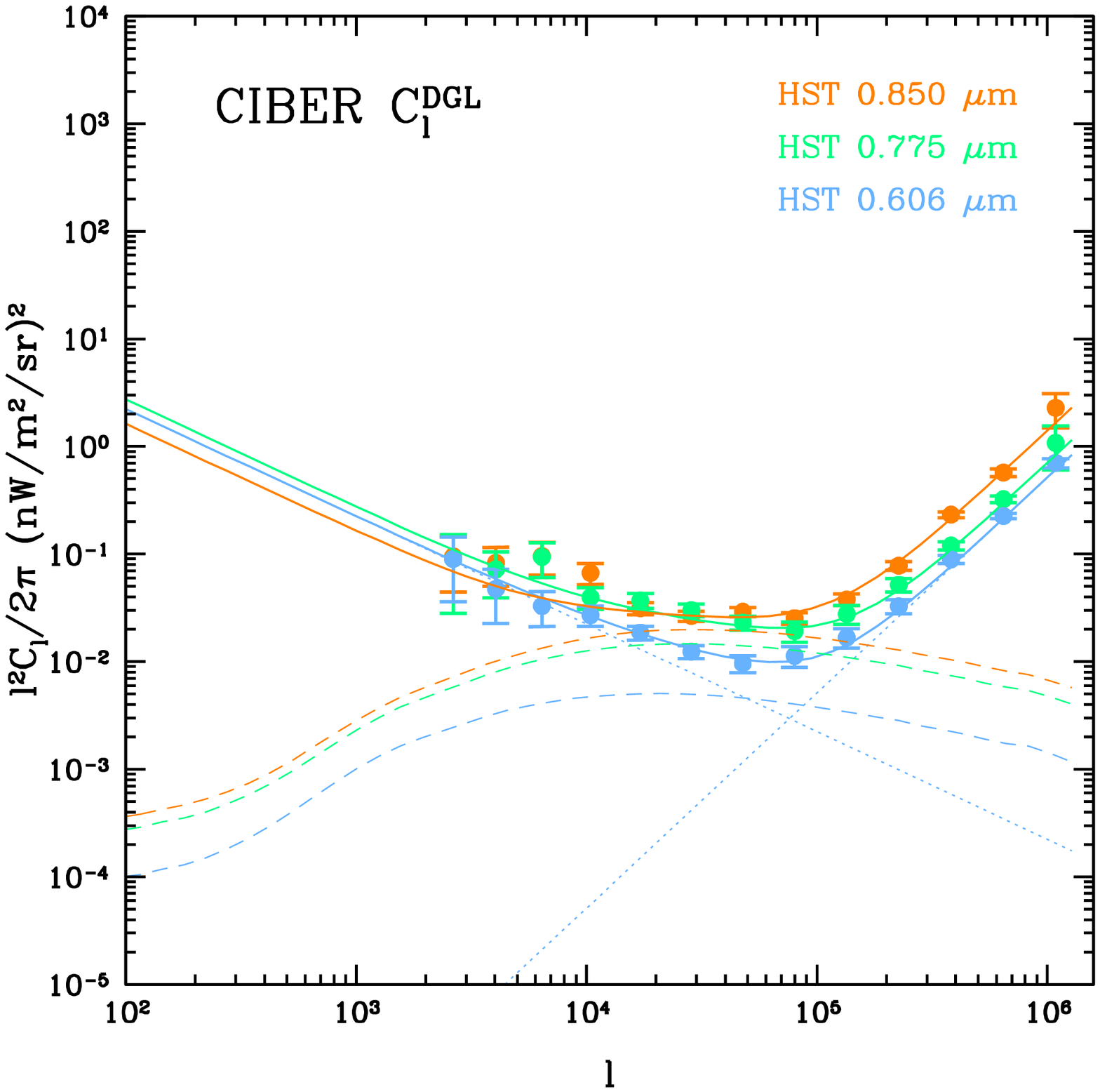}}
}

\epsscale{1.0}
\caption{\label{fig:Cl} The best-fit axion decay models for HST and CIBER anisotropy power spectra for the case with a mass distribution for axions given by Eq.~(\ref{eq:P_ma}). The solid, dashed and dash-dotted curves are the total, axion and high-$z$ power spectra, respectively. The shot-noise and $C_{\ell}^{\rm f}$ are shown in blue dotted lines for 0.606 $\mu$m and 1.6 $\mu$m as an example (we do not show them for other bands for clarity, but those terms are included in model fits).The long dashed curves are the low-$z$ power spectra from low-$z$ galaxies for the CIBER data. {\it Top:} The fitting results with free $C_{\ell}^{\rm f}$ for all bands. {\it Bottom:} The fitting results with CIBER $C_{\ell}^{\rm DGL}$, that is the foreground at large angular scales is the maximum allowed given the cross-correlation results between CIBER and IRAS (Zemcov et al. 2014), for 1.1, 1.25 and 1.6 $\mu$m. The fitting results for the case of single axion mass are quite similar.}
\end{figure*}

In order to fit the HST data, we consider four components including the axion decay model, the shot-noise term, the signal from high-$z$ faint galaxies during the epoch of reionization (EoR), and a power-law component that accounts for Galactic foregrounds, such as diffuse Galactic light (DGL). At small scales, the shot-noise dominates the power spectrum. Since it is scale-independent, its angular power spectrum is given by
\be
C_{\ell}^{\rm shot} = A_{\rm shot},
\ee
where $A_{\rm shot}$ is the shot-noise amplitude factor, which is a constant for a given observed wavelength. Note that we do not consider the contribution of low-redshift galaxies to HST measurements, since it just contributes to the shot-noise term at multipole moments of $\ell>2\times10^3$ and can be treated as shot-noise in these angular scales \citep{Helgason12}.

For the HST bands centered at 1.25 and 1.6~$\mu$m, the signal from high-$z$ faint galaxies can be an important component for the total signal. The bands of 1.25 and 1.6 $\mu$m can cover the high-$z$ signal from $8\lesssim z\lesssim 10.5$ and $10.5\lesssim z\lesssim 13$, respectively. We adopt an analytic model given by \cite{Cooray12a} to estimate the angular power spectrum of the high-$z$ galaxies. This model can provide the mean intensity and angular power spectra at different wavelengths with a few parameters, such as the star formation rate density (SFRD), the escape fraction of ionizing photons, and star formation efficiency. This model is also consistent with the observations of reionization history, optical depth of electron scattering and UV luminosity functions at high redshifts \citep{Cooray12a}. However, as claimed by \cite{Cooray12a}, this high-$z$ model cannot explain the anisotropies of nea-IR EBL since the amplitudes of high-z galaxy power spectra are at least two orders of magnitude smaller than the observations, and we need the other components to match the data. Here we add an amplitude factor in the model, and we have
\be
C_{\ell}^{\rm high-z} = A_{\rm high-z} C_{\ell,\rm model}^{\rm high-z},
\ee
where $A_{\rm high-z}$ is the amplitude factor which can be constrained by the data, and $C_{\ell,\rm model}^{\rm high-z}$ is the high-z power spectrum from \cite{Cooray12a}. The detection of $ A_{\rm high-z}$ from the HST data and the resulting implications in terms of the reionization model
are discussed in \cite{Mitchell-Wynne15}.

At large scales, the angular power spectrum is 
dominated by  $C_{\ell}^{\rm f}$ for foregrounds. We find it can be described by 
\be
C_{\ell}^{\rm f} = A_{\rm f} \ell^{-3},
\ee
where $A_{\rm f}$ is the amplitude factor for the foregrounds which can be fitted by the data. One possible and dominant component in $C_{\ell}^{\rm f}$ could be DGL, since the DGL  term is proportional to $\sim \ell^{-3}$ as shown in \cite{Zemcov14} and \cite{Mitchell-Wynne15}. 

Therefore, the model we use to fit the HST data at 1.25 and 1.6 $\mu$m is
\be
C_{\ell}^{\rm HST} = C_{\ell}^{\rm axion} + C_{\ell}^{\rm shot} + C_{\ell}^{\rm high-z}  + C_{\ell}^{\rm f}.
\ee
For the HST bands of 0.606, 0.775 and 0.850 $\mu$m, there is almost no high-$z$ signal received \citep{Mitchell-Wynne15}, so we fit the data by
\be
C_{\ell}^{\rm HST} = C_{\ell}^{\rm axion} + C_{\ell}^{\rm shot} + C_{\ell}^{\rm f}.
\ee

For the CIBER data at 1.1 and 1.6 $\mu$m, following \cite{Zemcov14}, we include the model of low-$z$ residual galaxies given by \cite{Helgason12} to calculate the low-$z$ contribution for the CIBER data.  This is due to the relatively shallow depth of the foreground mask in CIBER measurements
relative to HST fluctuations. The shallow depth is such that the clustering of residual galaxies make an appreciable contribution to CIBER
fluctuations. Then, we fit the CIBER data by
\be
C_{\ell}^{\rm CIBER} = C_{\ell}^{\rm axion} + C_{\ell}^{\rm high-z} + C_{\ell}^{\rm low-z} +  C_{\ell}^{\rm f}.
\ee
The $C_{\ell}^{\rm low-z}$ model is derived from the near-IR luminosity function at different bands. We add a scale factor $f_{\rm low-z}$ in the model to vary the low-$z$ angular power spectrum $C_{\ell}^{\rm low-z}$ in 1-$\sigma$ uncertainty, which means that $f_{\rm low-z}=0, 1$ and $0.5$ are for 1-$\sigma$ lower, upper and center values of $C_{\ell}^{\rm low-z}$, respectively. 
Note that the shot-noise term already has been included in $C_{\ell}^{\rm low-z}$ \citep{Zemcov14}. We use the same HST $C_{\ell}^{\rm high-z}$ and $C_{\ell}^{\rm f}$ at 1.25 and 1.6 $\mu$m to fit the CIBER $C_{\ell}^{\rm high-z}$ and $C_{\ell}^{\rm f}$ at 1.1 and 1.6 $\mu$m, respectively, since these two HST bands have similar bandwidths with the two CIBER bands. This could help us to fit the two terms, given that the HST and CIBER data have different scale coverages from $\ell=3\times10^2$ to $10^6$. Note that the $C_{\ell}^{\rm f}$ is fixed to be the upper limit of DGL power spectra $C_{\ell}^{\rm DGL}$  at 1.1 and 1.6 $\mu$m, measured in terms of the cross-correlation between CIBER and IRAS in \cite{Zemcov14}. In order to investigate this effect on the constraints of axion properties, we also explore the case with $C_{\ell}^{\rm f}$ fixed to be CIBER $C_{\ell}^{\rm DGL}$  at 1.1 and 1.6 $\mu$m. As we can see in the next section, the fitting results with CIBER $C_{\ell}^{\rm DGL}$ are similar to that with free $C_{\ell}^{\rm f}$ for the case of a single axion mass, and are consistent within 2-$\sigma$ C.L. for the case with an axion mass distribution, although the $\chi^2_{\rm min}$ of the fixed CIBER $C_{\ell}^{\rm DGL}$ case is much larger than that of the free $C_{\ell}^{\rm f}$ case.

In the model fitting process, we employ the Markov Chain Monte Carlo (MCMC) method to constrain the free parameters in the model. The Metropolis-Hastings algorithm is adopted to determine the probability of the acceptance of a new chain point \citep{Metropolis53, Hastings70}. The likelihood function can be estimated by $\mathcal{L}\propto {\rm exp}(-\chi^2/2)$, and $\chi^2$ is given by
\be
\chi^2 = \sum^{N_d}_{i=1} \frac{(C_{\ell}^{\rm obs}-C_{\ell}^{\rm th})^2}{\sigma_{\ell}^2},
\ee
where $N_d$ is the number of data points, $C_{\ell}^{\rm obs}$ and $C_{\ell}^{\rm th}$ are the angular power spectra from observational data and theory respectively, and $\sigma_{\ell}$ is the error of the data at $\ell$. In this work, we have the total $\chi^2=\chi^2_{\rm HST}+\chi^2_{\rm CIBER}$.

\begin{table*}[!t]
\begin{center}
\caption{\label{tab:best_fits}The best-fit values and 1-$\sigma$ errors of the free parameters in the models for the cases with an 
axion mass distribution and for a single mass with free $C_{\ell}^{\rm f}$ for all bands and CIBER $C_{\ell}^{\rm DGL}$ for 1.1, 1.25 and 1.6 $\mu\rm m$.}
\vspace{4mm}
\begin{tabular}{l | c | c | c | c}
\hline\hline
 Parameter & free $C_{\ell}^{\rm f}$ [$P(m_a)$] & CIBER $C_{\ell}^{\rm DGL}$ [$P(m_a)$] & free $C_{\ell}^{\rm f}$ ($m_a$)& CIBER $C_{\ell}^{\rm DGL}$ ($m_a$)\\
\hline
$m_0$ & $3.82^{+0.21}_{-0.20}$ &  $3.21^{+0.08}_{-0.07}$  & $-$ & $-$\\
$\alpha$ &  $1.67^{+0.47}_{-0.51}$  &  $1.66^{+0.21}_{-0.11}$  & $-$ & $-$\\
$\bar{m}_a$ or ${m}_a$ (eV) &  $4.39^{+0.18}_{-0.18}$  &  $3.91^{+0.12}_{-0.10}$  & $4.83^{+0.16}_{-0.14} $ & $5.03^{+0.10}_{-0.35}$\\
$\zeta$ &  $0.14^{+0.02}_{-0.05}$  &  $0.29^{+0.02}_{-0.03}$  & $0.021^{+0.003}_{-0.004} $  & $0.019^{+0.003}_{-0.001}$\\
${\rm log_{10}(A_{high-z})}$ &  $1.11^{+0.24}_{-0.31}$  &  $1.23^{+0.14}_{-0.21}$  & $1.28^{+0.18}_{-0.22} $  & $1.56^{+0.09}_{-0.10}$\\
$f_{\rm low-z}$ &  $0.47^{+0.03}_{-0.03}$  &  $0.52^{+0.03}_{-0.03}$  & $0.47^{+0.03}_{-0.03} $  & $0.52^{+0.03}_{-0.03}$\\
$A_{\rm f}^{0.606}$ &  $1.21^{+0.23}_{-0.22}\times 10^3$  &  $1.41^{+0.07}_{-0.06}\times 10^3$  &  $7.66^{+5.06}_{-1.01}\times 10^2 $  & $1.06^{+0.19}_{-0.22}\times10^3$\\
$A_{\rm f}^{0.775}$ &  $2.28^{+0.45}_{-0.62}\times 10^3$  &  $1.72^{+0.19}_{-0.15}\times 10^3$  & $2.67^{+0.24}_{-0.24}\times 10^3 $  & $2.75^{+0.30}_{-0.30}\times10^3$\\
$A_{\rm f}^{0.850}$ &  $2.20^{+0.31}_{-0.41}\times 10^3$  &  $1.02^{+0.29}_{-0.27}\times 10^3$  & $3.09^{+0.26}_{-0.45}\times 10^3 $  & $3.04^{+0.27}_{-0.28}\times10^3$\\
$A_{\rm f}^{1.1 \& 1.25}$ &  $4.79^{+0.39}_{-0.40}\times 10^4$  & $2.40\times10^3 $ & $4.24^{+0.96}_{-1.19} \times 10^4$  & $2.40\times10^3$\\
$A_{\rm f}^{1.6}$ &  $3.65^{+0.34}_{-0.34}\times 10^4$  & $1.28\times10^3$ & $3.67^{+0.50}_{-0.43}\times 10^4 $  & $1.28\times10^3$\\
$\rm C_{\ell,shot}^{0.606}$ $\rm [(nW/m)^2 sr^{-1}]$&  $3.44^{+0.11}_{-0.11} \times 10^{-12}$ & $3.22^{+0.09}_{-0.09}\times 10^{-12} $  &  $3.12^{+0.45}_{-0.18} \times 10^{-12} $  & $3.30^{+0.17}_{-0.16}\times 10^{-12} $\\
$\rm C_{\ell,shot}^{0.775}$ $\rm [(nW/m)^2 sr^{-1}]$&   $4.66^{+0.57}_{-0.34} \times 10^{-12}$ &  $4.41^{+0.18}_{-0.19}\times 10^{-12}$  & $4.90^{+0.20}_{-0.30} \times 10^{-12} $  & $4.88^{+0.36}_{-0.35}\times 10^{-12}$\\
$\rm C_{\ell,shot}^{0.850}$ $\rm [(nW/m)^2 sr^{-1}]$&  $8.56^{+0.37}_{-0.35} \times 10^{-12}$  &  $8.81^{+0.27}_{-0.33}\times 10^{-12}$  & $8.70^{+0.38}_{-0.68} \times 10^{-12} $  & $8.84^{+0.49}_{-0.71}\times 10^{-12} $\\
$\rm C_{\ell,shot}^{1.25}$ $\rm [(nW/m)^2 sr^{-1}]$&  $7.72^{+0.19}_{-0.22} \times 10^{-11} $ &  $7.93^{+0.19}_{-0.18}\times 10^{-11} $  & $7.77^{+0.21}_{-0.21} \times 10^{-11} $  & $7.34^{+0.23}_{-0.31}\times 10^{-11}$\\
$\rm C_{\ell,shot}^{1.6}$ $\rm [(nW/m)^2 sr^{-1}]$&  $7.58^{+0.10}_{-0.10} \times 10^{-11} $  &  $7.67^{+0.08}_{-0.09} \times 10^{-11}$  & $7.61^{+0.13}_{-0.13}\times 10^{-11} $  & $7.56^{+0.10}_{-0.11}\times 10^{-11}$\\
$\chi^2_{\rm min}$ & 264.5 & 617.3 & 289.2 & 705.0 \\
$\chi^2_{\rm red}\, (\chi^2_{\rm min}/{\rm dof})$ & 2.9 & 6.6 & 3.1 & 7.5 \\
\hline
\end{tabular}
\end{center}
\end{table*}

We assume the free parameters follow the flat prior probability distribution. For the case with a single axion mass, we assume $m_a\in(0,20)$, and we set $m_0\in(0,20)$ and $\alpha\in(0,5)$ for the case with an axion mass distribution. The other parameters and their ranges are as follow: $\zeta\in(0,1)$, ${\rm log_{10}}A_{\rm high-z}\in(-4,7)$, $f_{\rm low-z}\in(0,1)$. The factors of $A_{\rm f}$ and $C_{\ell}^{\rm shot}$ for five HST bands are also set as free parameters in our fitting process with the same ranges shown in \cite{Mitchell-Wynne15}. We generate twenty parallel MCMC chains, and about 100,000 chain points for each chain are collected after the convergence. We then merge all chains together and thin them to obtain about 10,000 points to illustrate the probability distribution functions (PDFs) of the free parameters. The details of our MCMC method can be found in \cite{Gong07}.

\section{Results}

In Figure~\ref{fig:Cl}, we show our best-fit results of the HST and CIBER data for the case with an axion mass distribution. 
The results for the case of a single axion mass are quite similar. The solid, dashed and dash-dotted curves are the total, axion and high-$z$ power spectra, respectively. The shot-noise term and $C_{\ell}^{\rm f}$ component are also shown for 0.606 $\mu$m as blue dotted lines as an example. The long dashed curves are the low-$z$ power spectra for the CIBER data at 1.1 and 1.6 $\mu$m. For 0.606, 0.775 and 0.850 $\mu$m, the anisotropies of axion emission (1-halo term) dominate the angular power spectra at $10^4<\ell<2\times10^5$, while the high-$z$ power spectra (1-halo term) contribute significantly to the power spectrum in the 1.25 and 1.6 $\mu$m bands at these scales, especially around $\ell=10^5$. At the smaller ($\ell>2\times10^5$) and larger scales ($\ell<10^4$), the angular power spectra are dominated by the shot-noise and $C_{\ell}^{\rm f}$, respectively. 

\begin{figure}[t]
\includegraphics[scale = 0.43]{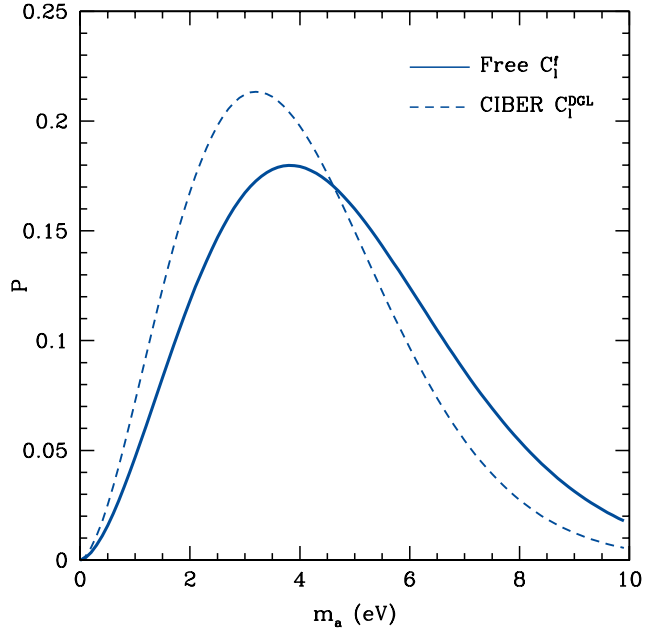}
\caption{\label{fig:ma_dis} The axion mass distributions $P(m_a)$ with the best-fit values of $m_0$ and $\alpha$. The blue solid line denotes the $P(m_a)$ with $m_0=3.82$ and $\alpha=1.67$ for the case of free $C_{\ell}^{\rm f}$, and the blue dashed line shows the result with $m_0=3.21$ and $\alpha=1.66$ for the case of CIBER $C_{\ell}^{\rm DGL}$.}
\end{figure}

For 1.1, 1.25 and 1.6 $\mu$m, we find the model with free $C_{\ell}^{\rm f}$ can fit the HST and CIBER data well with the $\chi^2_{\rm min}=264.5$ and reduced $\chi^2=2.9$. On the other hand,  we find $\chi^2_{\rm min}=617.3$ and the reduced $\chi^2=6.6$ for the case where we set $C_{\ell}^{\rm f}=C_{\ell}^{\rm DGL}$, that is the residual foreground which is the maximum level allowed by the CIBER cross-correlation with IRAS maps as described in Zemcov et al. (2014). For the case of a single axion mass, the results are similar, and we find $\chi^2_{\rm min}=289.2$ and the reduced $\chi^2=3.1$ for free $C_{\ell}^{\rm f}$, and $\chi^2_{\rm min}=705.0$ and reduced $\chi^2=8.0$ for CIBER $C_{\ell}^{\rm DGL}$. This indicates that there should be the other additional components, besides the measured CIBER $C_{\ell}^{\rm DGL}$, that contribute to this part of power spectrum. 

The best-fit values and 1-$\sigma$ errors of all free parameters in our models for all cases are shown in Table~\ref{tab:best_fits}. For the case of a 
single axion mass, we find the fitting results of the cases with free $C_{\ell}^{\rm f}$ and CIBER $C_{\ell}^{\rm DGL}$ are quite similar with each other.
The allowed parameter ranges of the axion model are consistent within 1-$\sigma$ confidence level (C.L.), although the
 $\chi^2_{\rm min}$ are much different between the two cases.

\begin{figure}[t]
\includegraphics[scale = 0.45]{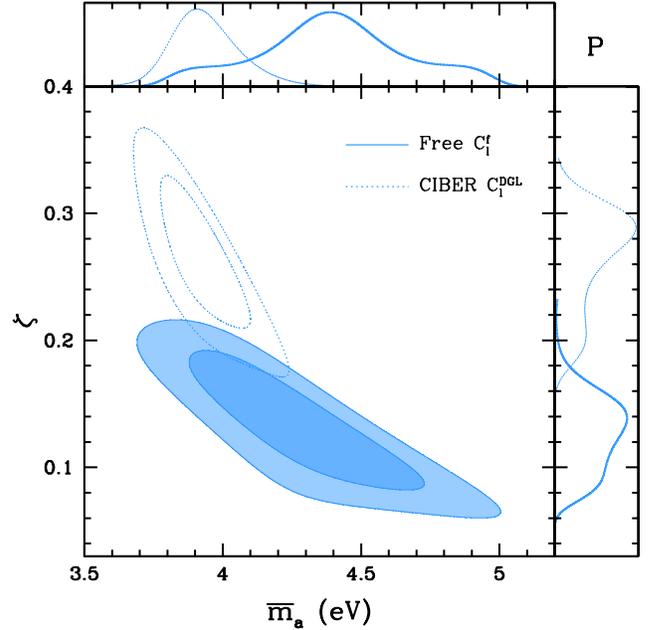}
\caption{\label{fig:ma_zeta} The contour maps of the mean axion mass $\bar{m}_a$ and the axion-photon coupling factor $\zeta$. The solid filled and dotted contours denote 1-$\sigma$ (68.3\%) and 2-$\sigma$ (95.5\%) C.L. of the constraint results for the cases of free $C_{\ell}^{\rm f}$ and CIBER $C_{\ell}^{\rm DGL}$, respectively. The corresponding 1-D marginalized PDFs of $\bar{m}_a$ and $\zeta$ are also shown in blue curves.}
\end{figure}

For the case of an axion mass distribution, we show the $P(m_a)$ with the best-fit values of $m_0$ and $\alpha$ in Figure~\ref{fig:ma_dis}. The blue solid and dashed lines are for the cases of free $C_{\ell}^{\rm f}$ and CIBER $C_{\ell}^{\rm DGL}$, respectively. We find they have the similar shapes with the peaks located around $3.5$ eV and wide distributions extending beyond 10 eV. The mean axion mass from these two $P(m_a)$ are $\bar{m}_a=4.45$ and $3.90$ for the free $C_{\ell}^{\rm f}$ case and the CIBER $C_{\ell}^{\rm DGL}$ case, respectively. Also, we derive the probability distribution for the mean axion mass $\bar{m}_a$ from the MCMC chains by integrating over the $P(m_a)$ calculated by each chain point with $m_0$ and $\zeta$.  We show the 2-D contour maps of $\bar{m}_a$ vs. $\zeta$ for both the cases of free $C_{\ell}^{\rm f}$ and CIBER $C_{\ell}^{\rm DGL}$ in Figure~\ref{fig:ma_zeta}. The 1-D marginalized PDFs for $\bar{m}_a$ and $\zeta$ are also shown in blue curves. We find that the constraint results for these two cases are different but consistent within 2-$\sigma$ C.L., and the best-fit values and 1-$\sigma$ errors are $\bar{m}_a=4.39^{+0.18}_{-0.18}$ and $\zeta=0.14^{+0.02}_{-0.05}$ for the free $C_{\ell}^{\rm f}$ case, and $\bar{m}_a=3.91^{+0.12}_{-0.10}$ and $\zeta=0.29^{+0.02}_{-0.03}$ for the CIBER $C_{\ell}^{\rm DGL}$. These values are also in a good agreement with the values of $\bar{m}_a$ derived from the $P(m_a)$ shown in Figure~\ref{fig:ma_dis}, which are obtained by the best-fit values of $m_0$ and $\alpha$.

Also, we find our fitting results of $A_{\rm high-z}$, $f_{\rm low-z}$, $A_{\rm f}$ and $C_{\ell}^{\rm shot}$ are similar to the results of \cite{Mitchell-Wynne15}. The axion decay model can explain the EBL fluctuation as equally well as the intrahalo light (IHL) model by comparing the $\chi^2_{\rm min}$with the IHL model, especially for the free $C_{\ell}^{\rm f}$ case \citep{Zemcov14,Mitchell-Wynne15}. Just based on a statistical comparison of the data
and model fits, we cannot distinguish between the scenarios involving IHL and axion decays to explain the clustering excess seen in 
the intensity fluctuations of the IR background.

\begin{figure}[t]
\includegraphics[scale = 0.43]{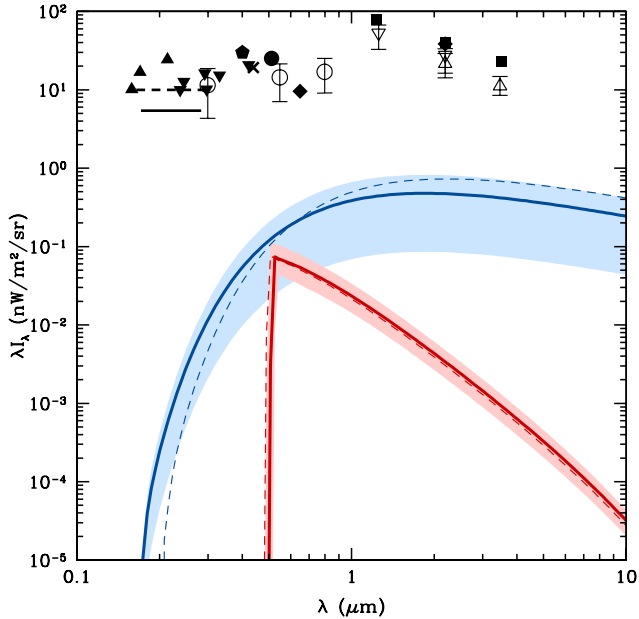}
\caption{\label{fig:Iv} The axion-photon decay intensity spectrum with the best-fits of $m_a$, $m_0$, $\alpha$ and $\zeta$. The blue and red curves denote the $\lambda I_{\lambda}$ for axion mass distribution and single mass cases, respectively. The solid and dashed curves are for the free $C_{\ell}^{\rm f}$ case and the CIBER $C_{\ell}^{\rm DGL}$, respectively. The shaded regions are the 1$\sigma$ uncertainties for the relevant curves. For comparison, we
show a subset of the absolute intensity measurements from the literature \citep{Overduin04}. The total intensity of the axion model is at most 1 nW m$^{-2}$ sr$^{-1}$, which is roughly a factor of 10-20 smaller than
the total galaxy contribution.}
\end{figure}

\begin{figure*}[htpb]
\epsscale{1.9}
\centerline{
\resizebox{!}{!}{\includegraphics[scale=0.45]{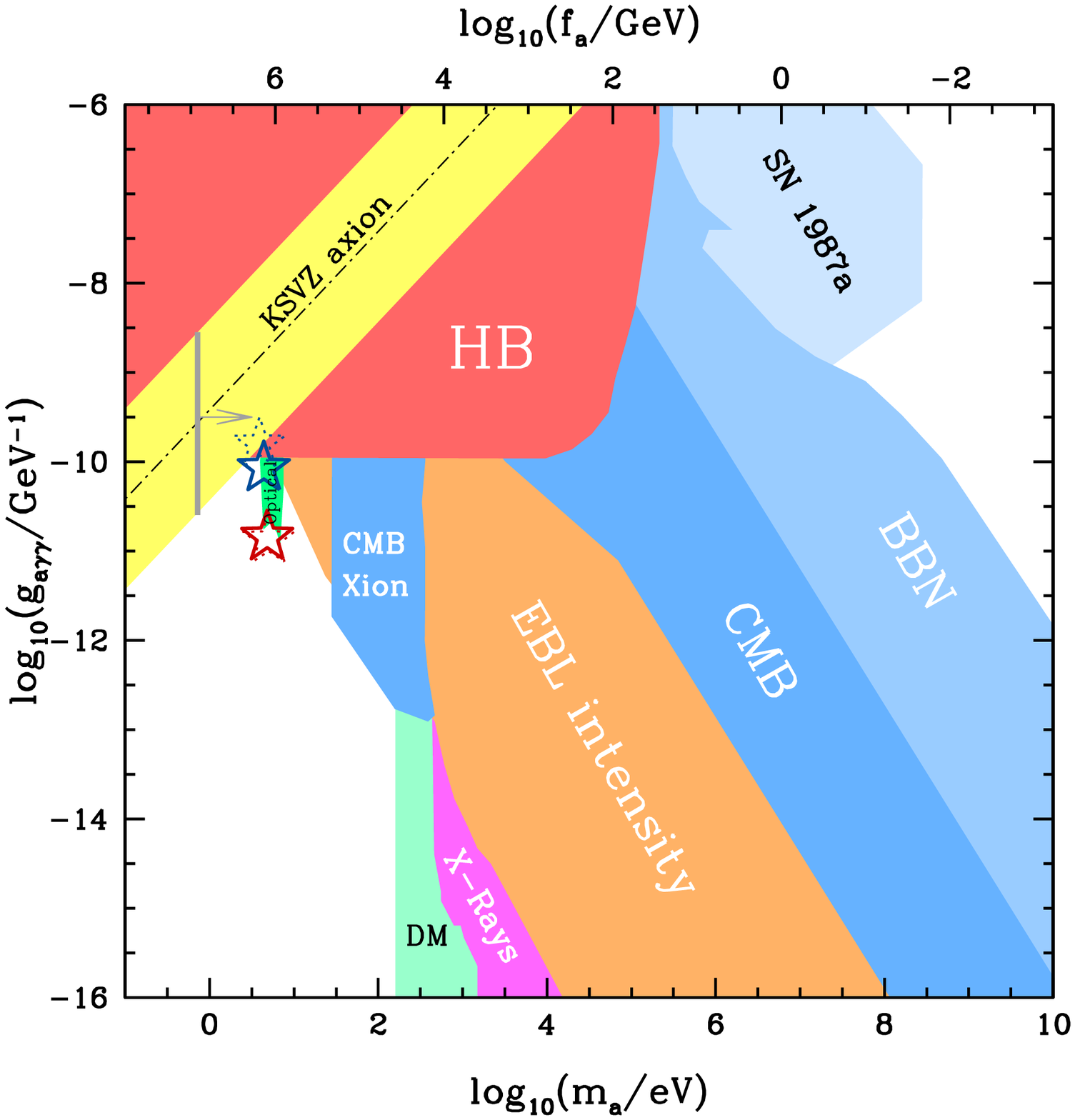}}
\resizebox{!}{!}{\includegraphics[scale=0.423]{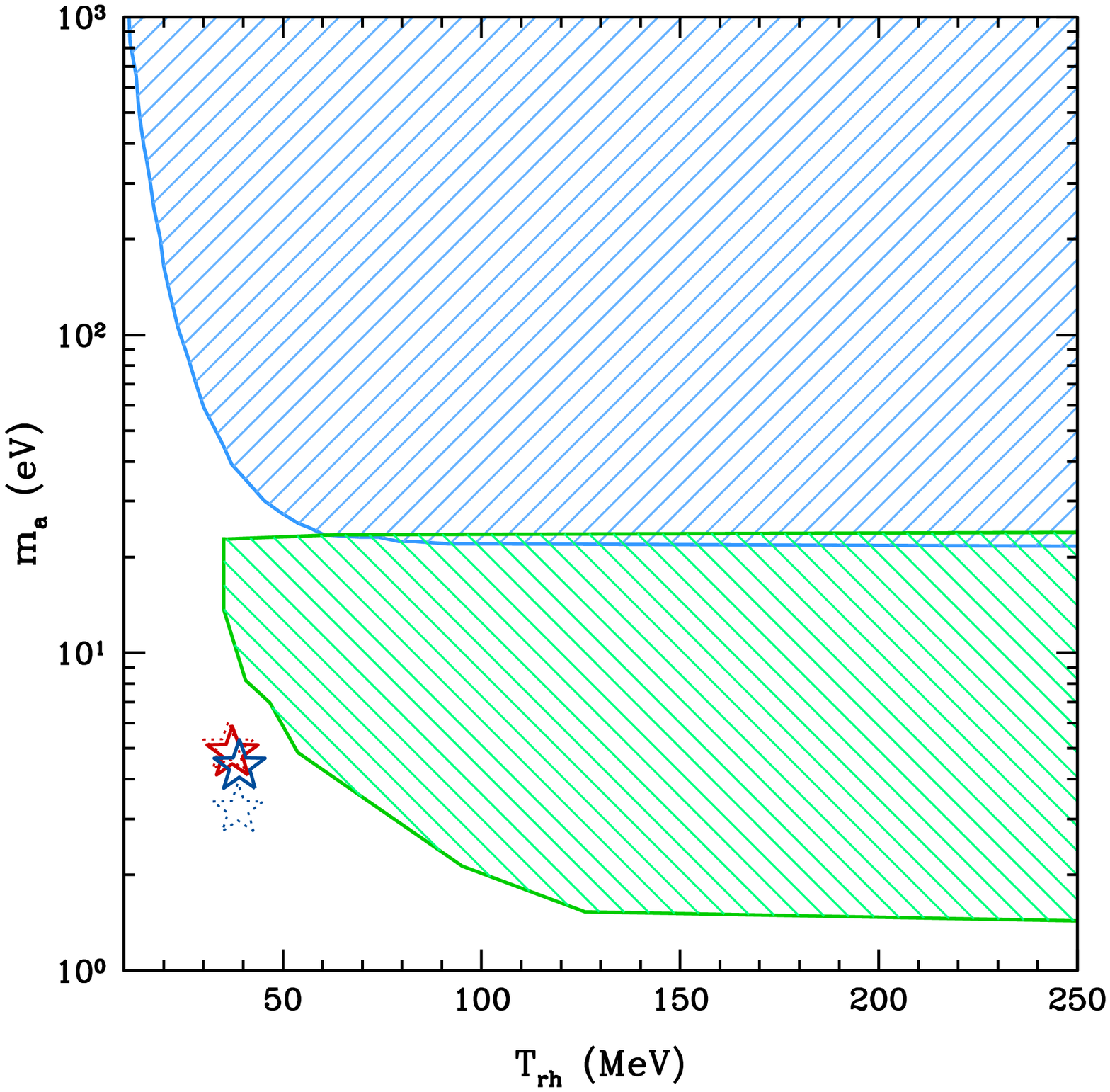}}
}
\epsscale{1.0}
\caption{\label{fig:comp} The comparisons of our best-fit axion model with the other measurements and estimations. Our best-fit axion models are shown in stars. The blue and red stars are for the axion mass distribution case and single mass case, respectively. The solid and dotted stars denote the free $C_{\ell}^{\rm f}$ case and the CIBER $C_{\ell}^{\rm DGL}$ case, respectively. $Left$ $panel$: the $m_a$ vs. $g_{a\gamma\gamma}$ diagram with the excluded regions from different measurements \citep{Grin07,Hannestad10,Cadamuro12}. $Right$ $panel$: the $m_a$ vs. $T_{\rm rh}$ diagram with the excluded regions given by \cite{Grin08}. }
\end{figure*}

In Figure~\ref{fig:Iv}, by using Eq.~(\ref{eq:Il}), we calculate the mean intensity spectra of the axion-photon decay with the best-fit values of $m_0$, $\alpha$, $m_a$ and $\zeta$ for the axion mass distribution case (blue curves) and single mass case (red curves). The solid and dashed curves are for the free $C_{\ell}^{\rm f}$ case and the CIBER $C_{\ell}^{\rm DGL}$ case, respectively. The shaded regions denote the 1-$\sigma$ uncertainties for corresponding curves. The observed data are also shown for comparison \citep{Overduin04}. As can be seen, the intensity spectra of the free $C_{\ell}^{\rm f}$ and CIBER $C_{\ell}^{\rm DGL}$ cases are consistent within 1-$\sigma$ C.L., although the fitting results of $m_0$ and $\alpha$ are not in very good agreement for these two cases (see Table~\ref{tab:best_fits}).  We find that our best-fit mean intensity spectra are much smaller than the measurements, by two orders magnitude at least. On one hand, the observed data can be overestimated by underestimating the mean intensity of the foreground contamination from Galactic diffuse light and zodiacal light. On the other hand, it indicates that the axion-photon decay only cannot provide enough intensity on the total near-IR intensity, although it could offer good interpretation for the excess of the near-IR anisotropies. The main contribution of the near-IR mean intensity might come from the low-redshift ($z<6$) galaxy emission (e.g. Gong et al. 2013).

We also compare our results with the other measurements and estimations in Figure~\ref{fig:comp}. In the left panel, we show the $m_a$ vs. $g_{a\gamma\gamma}$ diagram with excluded regions from different measurements, such as the constrains from the evolution of horizontal branch stars (HB), the duration of neutrino pulse of SN 1987a, big bang nucleosynthesis (BBN), cosmic microwave background (CMB), mean EBL intensity, dark matter (DM), X-rays, and optical observations \citep{Grin07,Cadamuro12}. We also show the bound from cosmological constraints given by the axion smooth effects on the power spectra of CMB and LSS (as a gray vertical line with an arrow) \citep{Hannestad10}. This bound excludes $m_a\gtrsim1$ eV along the KSVZ model (yellow region). In the right panel, the $m_a$ vs. $T_{\rm rh}$ with excluded regions are shown \citep{Grin08}. Here $T_{\rm rh}$ is the reheating temperature. The higher blue hatched region is excluded by the constraint $\Omega_ah^2<0.135$ assuming all dark matter is composed of axions. The lower green region shows the constraints from WMAP1 and SDSS data. 

We show our best-fit models in blue and red stars to denote the cases of axion mass distribution and single mass cases, respectively. The results of the free $C_{\ell}^{\rm f}$ case and CIBER $C_{\ell}^{\rm DGL}$ case are shown by solid and dotted stars, respectively. In comparison, we convert the best-fits of $m_a$ (or $\bar{m}_a$)  and $\zeta$ to $g_{a\gamma\gamma}$ using Eq. (\ref{eq:zeta_gamma}), and estimate $T_{\rm rh}$ via the relation of $\Omega_a$-$T_{\rm rh}$ given by \cite{Grin08}, with $\Omega_a$ obtained by Eq. (\ref{eq:Omega_a}) using the best-fit of $m_a$ or $\bar{m}_a$. As can be seen, the results of the single axion mass case are consistent with all current measurements. For the case of an axion mass distribution, the results stay safely outside the excluded regions of $m_a$ vs. $T_{\rm rh}$, but they have some tensions with the HB stars and optical observations. Fortunately, however, we find the constraint result of the free $C_{\ell}^{\rm f}$ case with $\bar{m}_a\lesssim 4$ eV is still consistent with the other observations (see the filled contours in Figure~\ref{fig:ma_zeta}), while the result of the CIBER $C_{\ell}^{\rm DGL}$ case (blue dotted star) is almost discarded by evolution of HB stars and LSS observations. Also, we need to note that the reheating temperature $T_{\rm rh}$ required by our model is only about 40 MeV, which is available in the low-temperature-reheating scenario \citep{Grin08}, it is however lower than most of theoretical expectations that are based on Big Bang nucleosynthesis constraints.

\section{Summary and discussion}

In this paper, we discuss the axion-photon decay as a potential origin of the anisotropy of the near-IR intensity fluctuations. According to the theoretical prediction, the axion particle can decay into two photons with wavelengths in optical and near-IR bands if the axion mass is within $\sim1-10$ eV. This provides a possible solution for the excess of the clustered anisotropies at near-IR wavelengths as seen in measurements from {\it Spitzer}, CIBER, {\it Hubble} and AKARI \citep{Kashlinsky05,Kashlinsky07,Matsumoto11,Kashlinsky12,Cooray12b,Zemcov14,Mitchell-Wynne15}. We calculate the luminosity of the axion decay at wavelength $\lambda$ in dark matter halos with mass $M$, and assume the single axion mass case and the mass distribution case in the estimation. With the help of the halo model, we estimate the mean intensity spectrum and the angular power spectrum for the axion-photon decay.

In order to compare with the observational measurements on the anisotropy power spectrum, 
we  adopt the MCMC method to fit our axion model with the data from all seven near-IR bands of HST and CIBER. We find our models for axion decay
 can fit the data  with equal statistical accuracy as the existing IHL model \citep{Cooray12b, Zemcov14}. The best-fits of the axion mass and photon coupling constant are $m_a=4.83$ eV and $\zeta=0.02$ for the single mass case, and the mean axion mass $\bar{m}_a=4.39$ and $\zeta=0.14$ for the case where we invoke a mass distribution. The single mass case, however, is ruled out by the lack
of cross-correlation of fluctuations between different wavelengths and the preferred model involves an axion mass distribution following
Eq.~(\ref{eq:P_ma}) with parameters given by the best-fit model.

We also compare the total mean intensity spectra of the axion decay  and find that the axion decay is producing at most a 1 nW m$^{-2}$ sr$^{-1}$, despite explaining the fluctuation measurements with HST and CIBER. Finally, we compare the best-fit values of $m_a$ (or $\bar{m}_a$) and $\zeta$ with the bounds given by different measurements and estimations, and find our models basically are not conflicting with these bounds, although there are some tensions between the constraints from the axion mass distribution case and the HB stellar evolution and optical observations. Therefore, our model can provide reasonable constraints on the axion properties, and can offer a possible explanation on the excess of anisotropies of the near-IR EBL.

We should also note that, beside the axion-photon decay, there are other possible models to explain the excess of the anisotropies of the near-IR EBL, such as the IHL from diffuse stars in dark matter halos \citep{Cooray12b,Zemcov14} and DCBHs formed at early Universe \citep{Yue13}. Hence, the constraints of the axion mass given in this work should be taken as an upper limit, since we assume the axion decay alone to provide the contribution to the EBL power spectra without considering the contributions from the models mentioned above. The axion mass and coupling strength can be smaller if including  other possible signals.

\begin{acknowledgments}
YG acknowledges the support of Bairen program from the National Astronomical Observatories, Chinese Academy of Sciences. YG and AC acknowledge the supports from NSF CAREER AST-0645427 and AST-1313319. XLC acknowledges the support of the MoST 863 program grant 2012AA121701, pilot B grant XDB09020301, and the NSFC grants 11373030. We thank an anonymous referee for helpful comments.
\end{acknowledgments}

\appendix

Besides the data from HST and CIBER discussed in the main paper, we also include the data from {\it Spitzer} fluctuation measurements at 3.6 $\mu$m given by \cite{Cooray12b} to perform a joint model fit with HST and CIBER. We take the similar model as the HST case to calculate the angular power spectrum $C_{\ell}^{\rm Spitzer}$ at 3.6 $\mu$m such that
\be
C_{\ell}^{\rm Spitzer} = C_{\ell}^{\rm axion} + C_{\ell}^{\rm shot} + C_{\ell}^{\rm high-z}  + C_{\ell}^{\rm f}.
\ee
We use the same axion and high-z faint galaxy model for all three datasets to estimate $C_{\ell}^{\rm axion}$ and $C_{\ell}^{\rm high-z}$, and use two parameters $A^{\rm 3.6}_{\rm shot}$ and $A^{\rm 3.6}_{\rm f}$ to estimate $C_{\ell}^{\rm shot}$ and $C_{\ell}^{\rm f}$ for Spitzer data, respectively.

\begin{figure*}[htpb]
\begin{center}
\includegraphics[scale = 0.5]{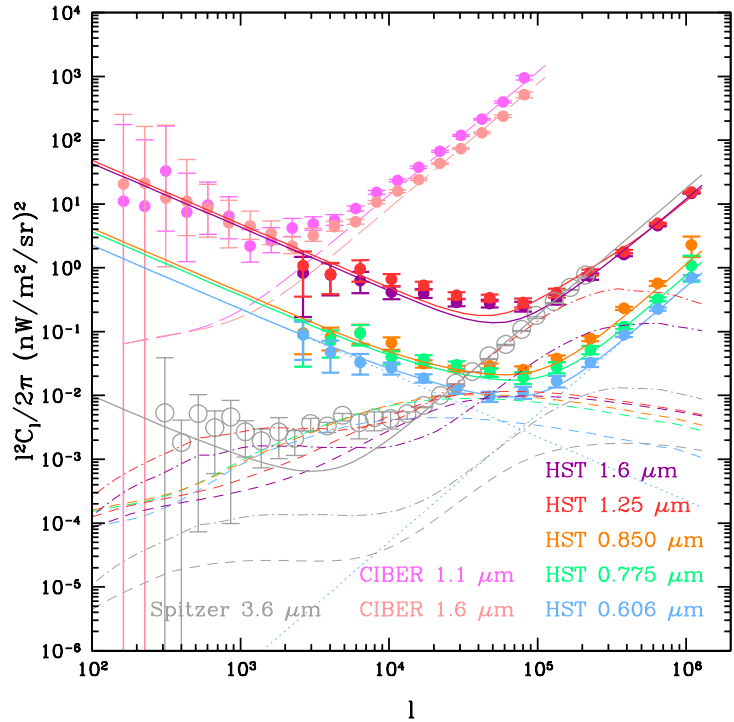}
\caption{\label{fig:Cl_Spi} The fitting results after including the data from {\it Spitzer} at 3.6 $\mu$m for the axion mass distribution case. The gray solid, dashed and dash-dotted curves are the total, axion decay and high-z galaxy power spectra at 3.6 $\mu$m, respectively. }
\end{center}
\end{figure*}

\begin{table}[!t]
\caption{\label{tab:best_fits_Spi}The best-fit values and 1-$\sigma$ errors of the free parameters in the model for the axion mass distribution case with HST, CIBER and Spitzer data.}
\vspace{4mm}
\begin{center}
\begin{tabular}{l | c }
\hline\hline
 Parameter & free $C_{\ell}^{\rm f}$ [$P(m_a)$] \\
\hline
$m_0$ &  $4.40^{+0.21}_{-0.22}$\\
$\alpha$ & $3.10^{+0.51}_{-0.66}$ \\
$\bar{m}_a$ & $4.53^{+0.29}_{-0.30}$\\
$\zeta$ &  $0.09^{+0.01}_{-0.01}$\\
${\rm log_{10}(A_{high-z})}$ &  $1.62^{+0.14}_{-0.16}$\\
$f_{\rm low-z}$ & $0.47^{+0.05}_{-0.04}$ \\
$A_{\rm f}^{0.606}$ & $1.42^{+0.24}_{-0.22} \times 10^3$\\
$A_{\rm f}^{0.775}$ & $2.30^{+0.31}_{-0.35} \times 10^3$ \\
$A_{\rm f}^{0.850}$ & $2.61^{+0.45}_{-0.47} \times 10^3$\\
$A_{\rm f}^{1.1 \& 1.25}$ & $3.04^{+0.44}_{-0.50} \times 10^4$\\
$A_{\rm f}^{1.6}$ & $2.73^{+1.37}_{-0.39} \times 10^4$\\
$A_{\rm f}^{3.6}$ & $6.19^{+1.09}_{-1.30}$\\
$\rm C_{\ell,shot}^{0.606}$ $\rm [(nW/m)^2 sr^{-1}]$& $3.21^{+0.17}_{-0.21}\times 10^{-12}$ \\
$\rm C_{\ell,shot}^{0.775}$ $\rm [(nW/m)^2 sr^{-1}]$&  $4.75^{+0.31}_{-0.50}\times 10^{-12}$ \\
$\rm C_{\ell,shot}^{0.850}$ $\rm [(nW/m)^2 sr^{-1}]$& $8.32^{+0.24}_{-1.26}\times 10^{-12}$ \\
$\rm C_{\ell,shot}^{1.25}$ $\rm [(nW/m)^2 sr^{-1}]$& $6.88^{+0.34}_{-0.24}\times 10^{-11}$ \\
$\rm C_{\ell,shot}^{1.6}$ $\rm [(nW/m)^2 sr^{-1}]$& $7.49^{+0.15}_{-0.16}\times 10^{-11}$ \\
$\rm C_{\ell,shot}^{3.6}$ $\rm [(nW/m)^2 sr^{-1}]$& $11.09^{+0.16}_{-0.18}\times 10^{-11}$\\
$\chi^2_{\rm min}$ & 372.6\\
$\chi^2_{\rm red}\, (\chi^2_{\rm min}/{\rm dof})$ & 3.2\\
\hline
\end{tabular}
\end{center}
\end{table}

In Figure~\ref{fig:Cl_Spi}, we show the fitting results for all there datasets including {\it Spitzer} at 3.6 $\mu$m for the case involving an axion mass distribution. The gray solid, dashed and dash-dotted curves are the best-fit results for the total, axion decay and high-z galaxy power spectra at 3.6 $\mu$m, respectively. The best-fit values and 1-$\sigma$ errors of the free parameters are shown in Table \ref{tab:best_fits_Spi}. We find that the fitting results of the free parameters are similar to the results from HST+CIBER data discussed in the main paper (second column of Table \ref{tab:best_fits}). Although the best-fits of $m_0$ and $\alpha$ are different for HST+CIBER and HST+CIBER+Spitzer, the derived mean axion mass $\bar{m}_a$ are consistent in 1-$\sigma$ confidence level.

We also find that the reduced $\chi^2$ is 3.2, which is comparable with the result of HST+CIBER case shown in Table \ref{tab:best_fits}. While the model  gives overall good fits for the HST and CIBER data, we find issues with the model fit to {\it Spitzer} fluctuations around $\ell=3\times 10^3$. The power spectrum of axion decay at 3.6 $\mu$m is relatively small. Hence, we interpret this result to imply that the axion decay alone cannot explain all of the near-infrared intensity fluctuations and other sources, such as IHL \citep{Cooray12b} and DCBHs \citep{Yue13}, are probably needed. However, since the reduced $\chi^2$ of HST+CIBER+Spitzer is comparable to that for HST+CIBER, the axion model still has the potential to fit the data of all bands. We will try to improve this model and explore the possibilities in the future work.

\end{document}